\documentclass[a4paper,twocolumn,fleqn]{article} 
\usepackage{amsmath}                 
\usepackage{txfonts}                   
\usepackage{mathptmx}                
\usepackage{cite}                    
\usepackage{mhchem}
\usepackage[pdftex]{graphicx}       
\usepackage{tabularx}
\usepackage{pfr_APiP2025}                 

\begin{document}

\title{A laboratory plasma experiment for X-ray astronomy using a compact electron beam ion trap (EBIT)}

\author{Yuki~AMANO\sup{1, 2}, Leo~HIRATA\sup{3, 2}, Moto~TOGAWA\sup{4, 5}, Hiromasa~SUZUKI\sup{6, 2}, Hiroyuki~A.~SAKAUE\sup{7}, Naoki~KIMURA\sup{7}, Nobuyuki~NAKAMURA\sup{8, 2}, Makoto~SAWADA\sup{9, 2}, Masaki~OURA\sup{10}, Jonas DANISCH\sup{4}, Joschka GOES\sup{4}, Marc BOTZ\sup{4}, Jos\'{e}~R.~CRESPO~L\'{O}PEZ-URRUTIA\sup{4}, and Hiroya~YAMAGUCHI\sup{1, 3, 2}}
\affiliation{
  \sup{1}The Institute of Space and Astronautical Science (ISAS), Japan Aerospace and Exploration Agency (JAXA), 3-1-1 Yoshinodai, Chuo-ku, Sagamihara 252-5210, Japan\\
  \sup{2}RIKEN Pioneering Research Institute, RIKEN, 2-1 Hirosawa, Wako, Saitama 351-0198, Japan\\
  \sup{3}Department of Physics, The University of Tokyo, 7-3-1 Hongo, Bunkyo-ku, Tokyo 113-0033, Japan\\
  \sup{4}Max-Planck-Institut f\"{u}r Kernphysik, Saupfercheckweg 1, 69117 Heidelberg, Germany\\
  \sup{5}European XFEL, Holzkoppel 4, 22869 Schenefeld, Germany\\
  \sup{6}Faculty of Engineering, University of Miyazaki, 1-1 Gakuen Kibanadai-nishi, Miyazaki 889-2192 Japan\\
  \sup{7}National Institute for Fusion Science, Toki, Gifu 509-5292, Japan\\
  \sup{8}Institute for Laser Science, The University of Electro-Communications,
Chofu, Tokyo 182-8585, Japan\\
  \sup{9}Department of Physics, Rikkyo University, 3-34-1 Nishi-ikebukuro, Toshima-ku, Tokyo 171-8501, Japan\\
  \sup{10}RIKEN SPring-8 Center, RIKEN, 1-1-1 Kouto, Sayo, Hyogo 679-5148, Japan\\
}

\date{}

\email{amano.yuki.t76@kyoto-u.jp}

\begin{abstract}
We present the basic performance and experimental results of an electron beam ion trap (JAXA-EBIT), newly introduced to the Japanese astronomical community. 
Accurate atomic data are indispensable for the reliable interpretation of high-resolution X-ray spectra of astrophysical plasmas.
The JAXA-EBIT generates highly charged ions under well-controlled laboratory conditions, providing experimental benchmarks for atomic data.
The JAXA-EBIT shows performance comparable to the Heidelberg compact EBIT through dielectronic recombination measurements of highly charged Ar ions.
Furthermore, we conducted resonant photoexcitation spectroscopy of highly charged ions using the soft X-ray beamline BL17SU at the synchrotron radiation facility SPring-8.
As a result, we successfully detected resonance transitions of He-like \ce{O^6+} and Ne-like \ce{Fe^16+}. 
These results demonstrate the capability of the JAXA-EBIT for precise measurement of atomic data and show that it serves as a powerful tool for advancing astrophysical research.

\end{abstract}



\AbstNum{O-16}
\CorrespondingAuthor{Yuki~AMANO}
\CorrespondingAffiliation{The Institute of Space and Astronautical Science, Japan Aerospace and Exploration Agency}
\PostalAddress{3-1-1 Yoshinodai, Chuo-ku, Sagamihara 252-5210, Japan}
\Tel{+81-70-3117-7331}
\Fax{+81-42-759-8455}



\maketitle

\begin{normalsize}

\newpage
\section{Introduction}
Over 95\% of the baryonic matter in the Universe exists in the form of high-temperature, X-ray emitting plasma. 
Therefore, X-ray observations of astronomical objects play a critical role in understanding the processes of structure formation and the chemical evolution of the Universe.
The X-Ray Imaging and Spectroscopy Mission (XRISM \cite{tashiro2018}), launched in September 2023, is equipped with a microcalorimeter Resolve \cite{ishisaki2022}, which provides high-resolution X-ray spectroscopy in the soft X-ray bandpass up to 12 keV with an energy resolution of $E/\Delta E \approx 1300$ at 6 keV.
This spectral resolution enables precise diagnostics of astrophysical plasmas, including detailed measurements of velocity structures \cite{xrism2025w49b, suzuki2025casa} and elemental abundances \cite{plucinsky2025} of the objects. 

Improvements in the energy resolution of X-ray spectrometers require plasma modeling based on more reliable atomic data.
The X-ray astronomy satellite Hitomi (ASTRO-H) \cite{takahashi2016astro}, launched in 2016, was equipped with the Soft X-ray Spectrometer (SXS) \cite{kelley2016astro}, a microcalorimeter with performance equivalent to XRISM Resolve.
Although the mission was short-lived due to a post-launch anomaly, Hitomi provided groundbreaking high-resolution X-ray spectra that highlighted the importance of accurate plasma modeling and atomic data.
Widely used plasma modeling frameworks in X-ray astronomy, including AtomDB/APEC \cite{smith2001, foster2012}, SPEX \cite{kaastra1996}, and CHIANTI \cite{dere1997, del2015}, rely on atomic data obtained from theoretical calculations.
Comparisons of Hitomi's observations with the theoretical spectra synthesized by these codes revealed that there are biases in parameter measurements, such as electron temperature and elemental abundances, due to non-negligible spectral modeling uncertainties caused by model assumptions and atomic data \cite{aharonian2018}.
To address this issue, it is necessary to establish experimental benchmarks to provide robust constraints on atomic data.

An electron beam ion trap (EBIT) is a device that produces highly charged ions using a magnetically compressed electron beam \cite{marrs1988measurement}.
EBITs have been employed not only in fundamental atomic physics \cite{shabaev1994hyperfine, vogel2013aspects}, such as testing relativistic and quantum electrodynamical effects, but also in applied studies relevant to fusion research \cite{kaufman1983magnetic, beiersdorfer2010iter, morita2013study} and astrophysical plasmas \cite{beiersdorfer2003laboratory, gharaibeh2011k, shah2016laboratory, gall2019ebit}.
Recent advanced EBITs, such as FLASH-EBIT \cite{epp2010x} and PolarX-EBIT \cite{micke2018}, can be combined with synchrotron radiation facilities, enabling resonant photoexcitation spectroscopy of highly charged ions.
This technique allows precise measurements of atomic data, such as transition energies and probabilities \cite{epp2007, kuhn2022}, which are important for astrophysical applications.

To address the growing demand for vast amounts of atomic data in X-ray astronomy, it is essential for X-ray astrophysics and atomic physics to be more closely connected than ever.
Therefore, we have introduced an EBIT to the Japanese astronomical community in collaboration with Japan Aerospace and Exploration Agency (JAXA) and Max-Planck-Institut f\"{u}r Kernphysik (MPIK). 
In this paper, we present the fundamental performance of the EBIT and experimental results obtained at the synchrotron radiation facility SPring-8.
Dielectronic recombination measurements demonstrate that our EBIT achieves performance comparable to the Heidelberg Compact EBIT at MPIK \cite{micke2018}.
Furthermore, we carried out high-resolution photoexcitation spectroscopy of highly charged ions at SPring-8, successfully measuring the resonance transitions of He-like \ce{O^6+} and Ne-like \ce{Fe^16+}.
This experimental method can be extended to comprehensive measurements of atomic data.
Throughout the paper, statistical errors are given at a 68\% confidence level.

\begin{figure*}
\begin{center}
\includegraphics[width=14.0cm]{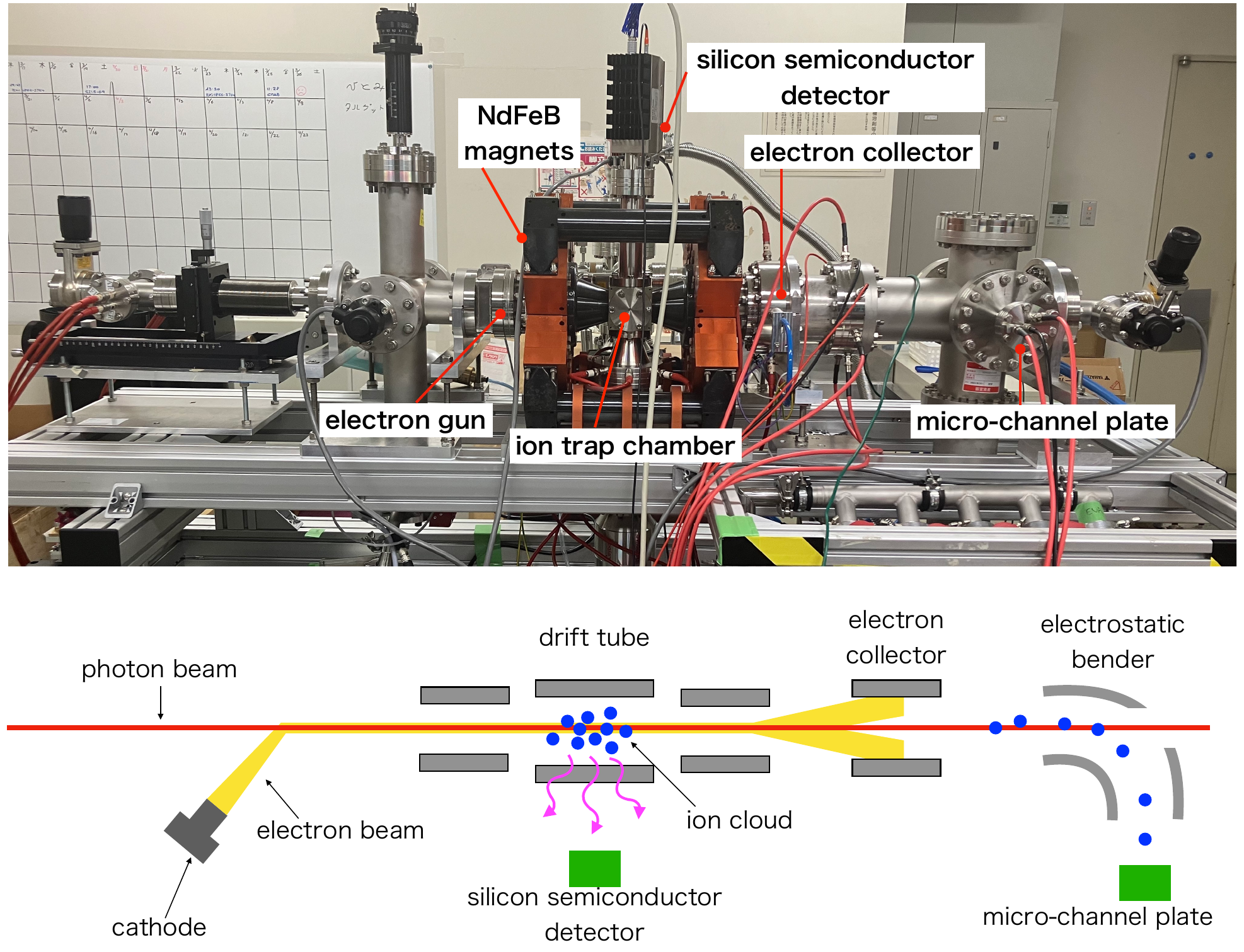} 
\end{center}
\caption{Photograph (upper panel) and schematic diagram (bottom panel) of the principal components of the JAXA-EBIT.
}
\label{fig:EBIT_view}
\end{figure*}

\section{Specification and performance of the EBIT}

Our EBIT (hereafter referred to as the JAXA-EBIT) was developed in collaboration with JAXA and MPIK.
The main specifications of the JAXA-EBIT are comparable to those of the Heidelberg Compact EBIT \cite{micke2018} developed by MPIK. 
Figure \ref{fig:EBIT_view} shows a photograph and schematic view of the principal components of the JAXA-EBIT.
Neutral precursor atoms of highly charged ions are injected as a molecular beam through a differential pumping system.
The injected atoms are subsequently ionized to the charge states of choice via successive electron impact ionization.
The ions are confined radially by the negative space-charge potential of the electron beam and axially by the potential differences applied to the drift tube.
The electron beam energy of the JAXA-EBIT is optimized over the range of 0.2--5.0 keV, allowing the production of Fe ions from Ar-like to He-like charge states.
The ions can be extracted by suddenly increasing the voltage on the central trap electrode in the drift tube \cite{kuhn2025}.
The ejected ions are detected by a micro-channel plate, and the ion charge-to-mass ratio can be resolved by time-of-flight measurement \cite{togawa2021}.
The JAXA-EBIT is equipped with an off-axis electron gun, which allows photon beam access to be coaxial with the electron beam without damaging the electron gun from synchrotron radiation (Figure \ref{fig:EBIT_view}). 
This specification allows precise measurement of atomic data by resonant photoexcitation spectroscopy \cite{togawa2024hanle, leutenegger2020, togawa2024high, steinbrugge2022}.

The JAXA-EBIT reproduces various ion--electron interactions, including collisional excitation, radiative recombination, and dielectronic recombination (DR) \cite{grilo2021} without injecting an external photon beam.
Investigating these processes is not only useful for obtaining atomic data but also for evaluating the performance of an EBIT.
We assessed the fundamental performance of the JAXA-EBIT, such as the monochromaticity and energy offset of the electron beam, through DR measurements.
DR is a resonant interaction between a free electron and an ion (Figure \ref{fig:DR}). 
A particularly important case for our purposes is the $KLL$ DR process, in which a free electron is captured into an open $L$-shell while a bound $K$-shell electron is simultaneously excited to the $L$-shell. 
Measurements of $KLL$ DR are especially useful for assessing electron gun performance, as the resonance energies and the corresponding X-ray emission intensities are well established \cite{takacs2015}.

We performed DR measurements by introducing Ar gas into the JAXA-EBIT.
We scanned the electron beam energy from 2210 to 2440 eV in 0.1 eV steps, while X-ray signals were recorded with a side-on-mounted silicon drift detector (SDD) for 10 seconds at each energy step.
We performed scans across the four energy bands shown in Figure \ref{fig:Ar_DR}, with the electron beam current at approximately 2 mA throughout the measurements. 
These energy ranges cover the DR resonance energies of highly charged Ar ions from C-like to He-like.
Figure \ref{fig:Ar_DR} (a) shows the scan results, where the horizontal axis corresponds to the electron beam energy, and the vertical axis to the pulse height of the SDD.
We find resonant enhancements of the X-ray yield at specific electron beam energies.
The projections of the detected signals within selected regions of interest in the detector pulse-height channels onto the electron beam energy axis are shown in Figure \ref{fig:Ar_DR} (b).
The obtained spectrum shows well-reproduced features of the previously reported $KLL$ DR spectrum of highly charged Ar ions \cite{biela2022}.

Since the Be-like DR resonance line at 2323 eV is well isolated from other DR lines, it provides a suitable reference for assessing the monochromaticity and energy offset of the electron beam.
We fit the line profile using a Gaussian plus linear function model. 
The resulting full width at half maximum (FWHM) and centroid energy of the Gaussian component are 2.33 $\pm$ 0.03 eV and 2323.38 $\pm$ 0.01 eV, respectively.
The measured width reflects broadening due to both ion motion and electron beam energy dispersion, and thus represents the effective energy resolution of the electron beam of the JAXA-EBIT.
The EBIT series is known to achieve excellent electron beam energy resolution by suppressing ion motion through strong confinement \cite{grell2024laboratory, grilo2025laboratory}.
In our measurements, the JAXA-EBIT achieved a resolving power of $E/\Delta E \approx 1000$ at 2.3 keV, broadly comparable to the reported performance of the Heidelberg Compact EBIT ($E/\Delta E \approx 1500$ at 5 keV \cite{micke2018}).
The observed centroid energy 2323.38 $\pm$ 0.01 eV is shifted by 13 eV from the value of 2310 eV obtained using the Flexible Atomic Code (FAC) \cite{gu2008, biela2022}.
This energy shift is roughly consistent with that expected from the space-charge potential measured in the Heidelberg Compact EBIT \cite{biela2022}.
These results confirm that the JAXA-EBIT demonstrates performance comparable to that of the Heidelberg Compact EBIT.

\begin{figure}
\begin{center}
\includegraphics[width=\linewidth]{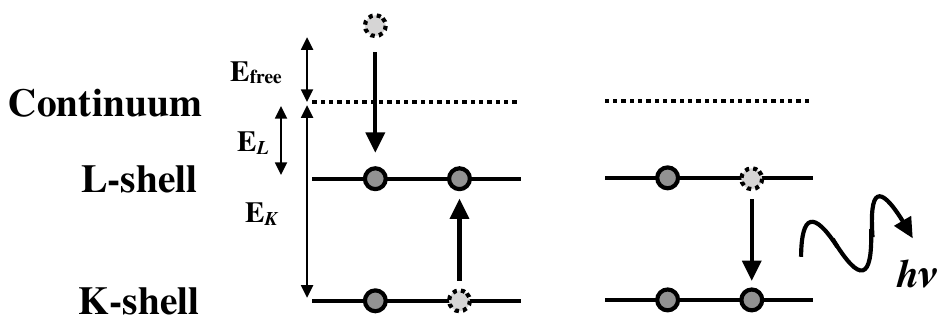} 
\end{center}
\caption{Schematic diagram of the $KLL$ dielectronic recombination of a He-like ion.
A free electron with kinetic energy $E_{\rm free}$ is captured into the $L$-shell with binding energy $E_{L}$, while a bound $K$-shell electron is simultaneously excited to the $L$-shell by the released energy $E_{\mathrm{free}} + E_{L}$.
The DR resonance occurs when the condition $E_{K} - E_{L}= E_{\mathrm{free}} + E_{L}$ is satisfied.
The resulting Li-like ion subsequently decays via K$\alpha$ X-ray emission.
}
\label{fig:DR}
\end{figure}

\begin{figure*}
\begin{center}
\includegraphics[width=14.5cm]{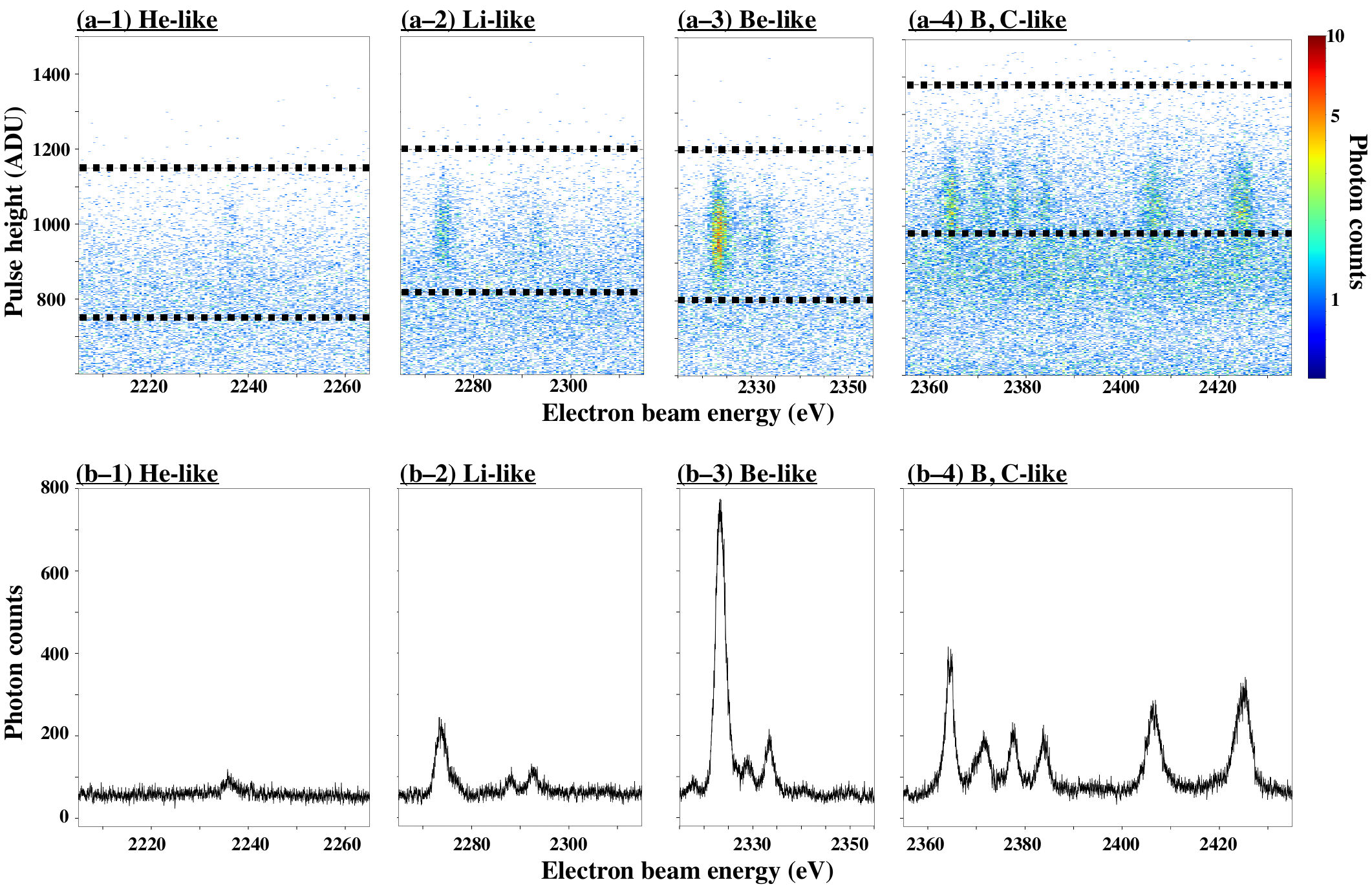} 
\end{center}
\caption{Results of dielectronic recombination measurements of highly charged Ar ions with the JAXA-EBIT.
(a) Count maps around the resonance energies of different charge states: (a--1) He-like, (a--2) Li-like, (a--3) Be-like, and (a--4) B- and C-like.
The horizontal axis corresponds to the electron beam energy, and the vertical axis to the pulse height of the X-ray detector.
(b) Projected spectra obtained by selecting events within the regions enclosed by the black dashed lines in panel (a) and projecting them onto the electron beam energy axis.
}
\label{fig:Ar_DR}
\end{figure*}

\section{Beam line experiment}

In June 2024, we conducted high-resolution photoexcitation spectroscopy of highly charged ions at the synchrotron radiation facility SPring-8.
In this experiment, we successfully measured the $L$-shell transition of Ne-like \ce{Fe^16+} and the $K$-shell transition of He-like \ce{O^6+}.
Ne-like \ce{Fe^16+} L-shell lines are prominent spectral features of astrophysical plasmas observed in the soft X-ray band below 1 keV \cite{pinto2015}.
Compared with Fe $K$-shell lines, the $L$-shell transitions provide richer photon statistics, allowing robust plasma diagnostics \cite{xu2002, beiersdorfer2018}.
For example, the intensity ratio between the ($3d$--$2p$) and ($3s$--$2p$) transitions serves as a probe of plasma opacity \cite{de2012, amano2020}, owing to the large difference in their oscillator strengths \cite{loch2005}.
However, the diagnostic power of this ratio is limited by discrepancies between observational results and theoretical predictions of the oscillator and collision strengths \cite{laming2000, beiersdorfer2002, beiersdorfer2004, shah2019}.
Resonant photoexcitation spectroscopy offers one of the most accurate experimental approaches for constraining oscillator strengths \cite{epp2007, rudolph2013, kuhn2022}.
We therefore carried out the beamline experiment to constrain the oscillator strengths ratio of the ($3d$--$2p$) transition, known as the 3C line ($[2p^53d_{3/2}]_{J=1} \rightarrow [2p^6]_{J=0}$), and the ($3s$--$2p$) transition, known as the 3G line ($[2p^53s_{1/2}]_{J=1} \rightarrow [2p^6]_{J=0}$).
A detailed description of the experimental methods and results is provided in \cite{hirata2025}.

The experiment was carried out at the soft X-ray beamline BL17SU \cite{ohashi2007, senba2007, tanaka2023} of SPring-8.
The JAXA-EBIT was connected to BL17SU through a differential pumping system, and a monochromatic photon beam was irradiated onto the highly charged ions trapped inside the EBIT.
The incident photon energy was calibrated before the beam time by conducting X-ray photoemission spectroscopy measurements of 4f electrons from the thin Au foil.
The photon beam energy was scanned around the expected resonance energies of each transition.
X-ray photons emitted from resonantly excited ions were recorded by a side-mounted SDD equipped with an Al filter, positioned perpendicular to the photon beam axis.

Figure \ref{fig:beamline_results} presents the results for the He-like \ce{O^6+} $2p$--$1s$ resonance transition line and the Ne-like \ce{Fe^16+} 3C and 3G lines.  
The well-known \ce{O^6+} He$\alpha$ line at 574 eV was used both as an energy reference for calibration and as an alignment check to ensure precise overlap of the photon beam with the trapped ion cloud.
By fitting the \ce{O^6+} He$\alpha$ spectrum with a Gaussian plus linear background model, the centroid energy was determined to be 573.776 $\pm$ 0.001 eV, revealing a centroid energy shift of $-0.17$ eV from the reference value of 573.95 eV adopted in AtomDB.
For the \ce{Fe^16+} $L$-shell transitions, the 3C line was successfully detected, whereas the 3G line, with a relatively small oscillator strength, was not clearly observed.
The 3C line fit provided a centroid energy of 825.658 $\pm$ 0.003 eV and a Gaussian area of 1.06 $\pm$ 0.06 counts s$^{-1}$ eV.
The centroid energy is shifted by -0.17 eV relative to the theoretical value of 825.83 eV \cite{loch2005}, consistent with the offset found for the \ce{O^6+} He$\alpha$ line.
Assuming that the centroid energy of the 3G line is shifted by 0.17 eV from the current experimental value of 727.07 eV \cite{shah2024high}, we obtained a 95 \% upper limit of the Gaussian area of the 3G line of 0.378 counts s$^{-1}$ eV.

We obtained centroid energies and Gaussian area by spectral fitting.
We constrain the centroid energy of the 3C line to 825.658 $\pm$ 0.003 eV, corresponding to a relative accuracy of $\Delta E/E \approx 4 \times 10^{-6}$.
However, both the 3C line and the He-like \ce{O^6+} resonance transition exhibited a systematic shift of $\approx 0.17$ eV from their theoretical values. 
This offset is considered to arise from systematic uncertainties in our experimental setup, such as the calibration of the synchrotron photon energy.
In our experiment, the calibration of the synchrotron photon energy was performed one month before the beamtime.
Recent high-precision measurements suppressed the systematic uncertainty to 10--15 meV by simultaneously tracking detailed photon energy fluctuations with a high-resolution photoelectron spectrometer \cite{shah2024Fe}.
Another reliable approach is to use well-established reference lines of highly charged ions confined within the EBIT, such as the He$\alpha$, He$\beta$, Ly$\alpha$, and Ly$\beta$ transitions.
Previous studies demonstrated that this method can calibrate the beam energy with a relative accuracy of $\Delta E/E \sim 10^{-6}$ \cite{leutenegger2020}.
The Gaussian area ratio of the 3C to 3G lines is directly proportional to their oscillator strength ratio \cite{bernitt2012}. 
After correcting the Gaussian areas for photon beam flux and exposure time, we constrained this ratio to an upper limit of $f_{3G}/f_{3C} < 0.322$ at the 95 \% confidence level.
Meanwhile, the limited signal-to-noise ratio prevented a clear detection of the weaker \ce{Fe^16+} 3G line.
The detection of such weak transitions can be achieved either by increasing the photon flux, for example, using next-generation beamlines, such as NanoTerasu (e.g., \cite{ohtsubo2022design, obara2025commissioning}) and SPring-8-II \cite{tanaka2024green}, or by reducing the experimental background through improvements in detector energy resolution.


\begin{figure*}
\begin{center}
\includegraphics[width=\linewidth]{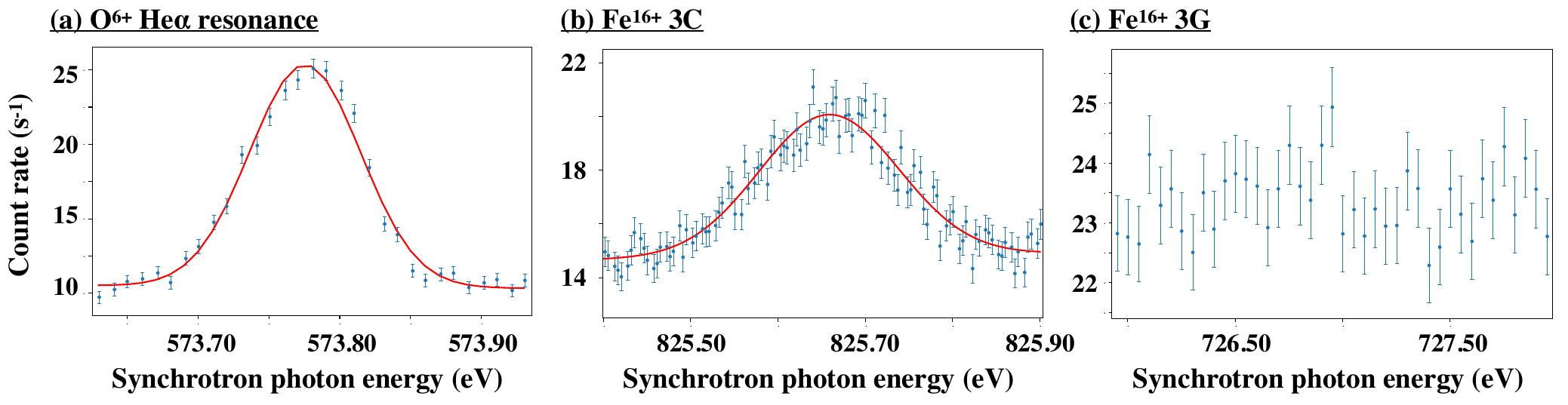} 
\end{center}
\caption{Results of photoexcitation spectroscopy at the soft X-ray beamline BL17SU of SPring-8.
(a) The \ce{O^6+} He$\alpha$ resonance line, (b) the \ce{Fe^16+} 3C transition, and (c) the \ce{Fe^16+} 3G transition.
The horizontal axis represents the incident photon energy of the synchrotron beam, and the vertical axis shows the count rate recorded by the silicon drift detector. 
The red solid lines indicate the best-fit models (Gaussian plus linear background).
Detailed descriptions of the experimental method and analysis procedures are provided in \cite{hirata2025}.
}
\label{fig:beamline_results}
\end{figure*}

\section{Future prospects}
%
%
The XRISM has observed various high-energy astrophysical phenomena, including supernova remnants \cite{suzuki2025casa, plucinsky2025, xrism2024n132d, xrism2025SagittariusAEast, vink2025, bamba2025}, active galactic nuclei \cite{audard2024xrismNGC4151, miller2025xrism, mehdipour2025delving}, galaxy clusters \cite{xrism2025Abell2029, xrism2025bulk, audard2025xrism}, and X-ray binaries \cite{audard2024xrismCygX3, mochizuki2024detection, tsujimoto2025outflowing, miura2025xrism}, and is continuously producing pioneering results.
At the same time, these observations have highlighted the urgent need for new atomic data.
For example, the transition energies and probabilities of K$\alpha$ and K$\beta$ inner-shell transitions of intermediately-charged ions (from Ne-like to Li-like) are crucial for interpreting the spectra of supernova remnants and X-ray binaries.
XRISM enabled the detection of X-ray emission from previously elusive odd-Z elements such as P, Cl, and K in supernova remnants (XRISM collaboration in prep.).
However, several of these lines cannot be clearly distinguished from the K$\beta$ transitions of Li-like Si and S, and uncertainties in the intensities of these K$\beta$ lines limit the reliability of abundance determinations.
Because the excited states of the K$\beta$ transition can decay not only through radiative transitions but also via Auger processes, measurements of the branching ratios between these de-excitation channels are essential.

The JAXA-EBIT is suited for such studies, as it allows simultaneous synchrotron radiation injection and ion extraction, enabling direct measurements of both radiative and Auger decay probabilities \cite{steinbrugge2015absolute}.
In addition, comprehensive atomic data measurements require beamlines covering a broad energy range.
A tender X-ray beamline, such as BL13U \cite{ohtsubo2022design, obara2025commissioning} at NanoTerasu, is ideal for probing the K-shell transitions of Si and S, while hard X-ray beamlines at SPring-8 are optimal for measurements of the Fe $K$-shell lines.
By combining the JAXA-EBIT and these state-of-the-art beamlines with XRISM observations, we can establish a robust framework for producing benchmark atomic data and gain new insights into both astrophysics and atomic physics.

\section{Conclusion}
We presented the performance and experimental results of the JAXA-EBIT.
Dielectronic recombination measurements confirmed that its performance is comparable to the Heidelberg Compact EBIT at MPIK.
We conducted the photoexcitation spectroscopy of highly charged ions using the JAXA-EBIT and the soft X-ray beamline BL17SU at SPring-8, successfully detecting the \ce{O^6+} He$\alpha$ and \ce{Fe^16+} 3C transitions.
These results provide experimental constraints on key atomic data, including transition energies and probabilities, which are indispensable for the interpretation of astrophysical X-ray spectra.
Extending this approach to a wider range of transitions with next-generation beamlines will yield benchmark atomic data that are essential for XRISM and forthcoming high-resolution X-ray missions.

\subsection*{Acknowledgments}
The authors thank Dr. Atsushi Takada and Dr. Takeshi Go Tsuru for their help in the experiment.
The synchrotron radiation experiments were performed at beamline BL17SU of SPring-8 with the approval of the RIKEN SPring-8 Center (proposal No. 20240064).
The authors deeply appreciate all the technical team members of BL17SU at SPring-8. 
This work is supported by JSPS/MEXT Scientific Research grant Nos. JP24K17106 (Y.A.), JP22H00158 (H.Y.), and JP23H01211 (H.Y.), NIFS Collaboration Research Program grant Nos. NIFS25KSPQ003, NIFS25KIIQ019.

\end{normalsize}


\bibliography{reference}{}
\bibliographystyle{unsrt}

\end{document}